\documentclass[reprint, superscriptaddress, aps, longbibliography, pre]{revtex4-2}

\usepackage{nicefrac, dsfont, amsmath, amsfonts, amssymb, amsthm, MnSymbol, mathrsfs, graphicx, comment, times, mathtools, blkarray, xcolor, bm, soul, physics, cancel, notes2bib, overpic}

\DeclarePairedDelimiterX{\KLDx}[2]{(}{)}{
  #1 \vert\vert #2
}
\newcommand{\KLD}{D\KLDx}

\graphicspath{{figures/}}

\begin{document}

\title{Thermodynamic efficiency of communication channels}

\author{Nahuel Freitas}
\email{nfreitas@df.uba.ar}
\affiliation{Universidad de Buenos Aires, Facultad de Ciencias Exactas y Naturales, Departamento de F\'isica. Buenos Aires, Argentina}

\author{Pedro E. Harunari}
\email{pedro.harunari@cnrs.fr}
\affiliation{Complex Systems and Statistical Mechanics, Department of Physics and Materials Science, University of Luxembourg, 30 Avenue des Hauts-Fourneaux, L-4362 Esch-sur-Alzette, Luxembourg}
\affiliation{Aix Marseille Université, CNRS, CINAM, Turing Center for Living Systems, 13288 Marseille, France}

\author{Massimiliano Esposito}
\email{massimiliano.esposito@uni.lu}
\affiliation{Complex Systems and Statistical Mechanics, Department of Physics and Materials Science, University of Luxembourg, 30 Avenue des Hauts-Fourneaux, L-4362 Esch-sur-Alzette, Luxembourg}

\begin{abstract}
We identify a broad class of communication channels that captures common physical constraints in both artificial and natural systems and derive bounds on their thermodynamic cost.
We find that the entropy production per channel use is bounded from below by the input–output mutual information, and their ratio---mutual information divided by entropy production---defines the thermodynamic efficiency. 
Unlike previous studies of energy-constrained communication channels, our analysis shows that thermodynamic costs must be assigned not only to the input symbols themselves, but also to transitions between successive symbols. 
As a result, maximizing thermodynamic efficiency favors a biased input that switches only rarely, rather than the capacity-achieving input. For the binary symmetric channel, this preference emerges through a pitchfork bifurcation that spontaneously breaks the symmetry of the channel. A minimal model of cellular sensing exhibits the same phenomenon.

\end{abstract}
\maketitle

\textit{Introduction}---
Energetic considerations have been central to information theory since its earliest developments. A paradigmatic example is given by the continuous Gaussian channel with additive white noise, for which the channel capacity depends explicitly on the power of both the signal and the noise, which is typically of thermal origin \cite{Hartley1928Jul, shannon1948mathematical, shannon2006communication, Cover2005Apr}. This suggested a fundamental connection between thermodynamic cost and communication capacity, that was for a long time taken for granted \cite{Landauer1996Jun}. However, as first pointed out by Landauer \cite{Landauer1996Jun}, the power entering the capacity formula for a Gaussian channel need not be dissipated in the transmission medium and therefore cannot be directly related to the entropy produced during transmission, which is the relevant thermodynamic cost. 
In fact, Landauer argued that communication imposes no fundamental thermodynamic cost. He showed that it is indeed possible to imagine communication protocols that are able to transmit information by employing carefully crafted time-dependent potentials, in such a way that the thermodynamic cost is negligible in the adiabatic limit~\cite{Landauer1996Jun, bennettNotesLandauersPrinciple2003, sagawaThermodynamicLogicalReversibilities2014}.

Yet, in many biological and artificial systems, communication carries a significant thermodynamic cost. The absence of a universal lower bound does not mean that costs are negligible or unimportant; rather, it suggests that the actual cost of any given communication scheme must be understood by examining the detailed physics of that scheme. 
This perspective was pursued by Bryant and Machta \cite{Bryant2023Aug}, who analyzed several biologically relevant modes of signaling — including electrical signaling via membrane depolarization through ion channels, diffusive signaling in two and three dimensions, and acoustic signaling. As the authors emphasize, the resulting cost bounds ``can only be obtained by making reference to the physics of the system; they cannot be extracted from information-theoretic considerations alone''.

In this Letter we take an approach that sits midway between the recognition that communication, as such, has no universal thermodynamic cost, and the conclusion that any meaningful bound must rely on the detailed physics of a specific communication mechanism. We identify a generic family of communication schemes that captures common physical constraints encountered in both artificial systems, such as electronic circuits~\cite{helmsStochasticThermodynamicBounds2025, chenOptimalControlBit2026}, and natural systems, such as biomolecular networks~\cite{Mattingly2021Dec, arunachalamInformationGainLimit2025}, while remaining amenable to a transparent thermodynamic analysis. Within this framework, we derive a lower bound on the average entropy produced per channel use in terms of the mutual information between input and output. We then define a thermodynamic efficiency for communication channels, upper bounded by unity, that directly measures how far a given channel operates from this bound. We show that the optimization of this efficiency, in general, does not coincide with the maximization of mutual information, and can lead to a non-trivial symmetry-breaking transition in the optimal input distribution. This phenomenon is illustrated in the paradigmatic example of the binary symmetric channel, which we analyze in detail. 

Notably, this symmetry breaking is absent from previous treatments of
energy-constrained communication~\cite{verdu1990capacity,
balasubramanian2001metabolically}, which assign an energy cost to each
transmitted symbol. Such per-symbol costs are physically sensible and are
recovered in our formalism, but they capture only part of the dissipation: a cost
must also be ascribed to the transitions or \emph{switches} between input symbols, and for
channels operating under detailed balance conditions this is the only source of dissipation
at all. It is precisely this switching cost that biases the efficiency-optimal input distribution away from the one compatible with the channel symmetries.

\textit{Information Theory}---A noisy, memoryless, and discrete communication channel is defined by an input alphabet $\mathcal{X}$, an output alphabet $\mathcal{Y}$, and the set of probabilities $\pi_{y \vert x}$ of transmitting a symbol $y \in \mathcal{Y}$ given an input $x \in \mathcal{X}$. Then, if each input symbol has a probability $\pi_x$, the output distribution is $\pi_y = \sum_x \pi_{y \vert x} \pi_x$. A measure of the information shared between the input and the output is given by the mutual information
\begin{equation}
    I = \sum_{x} \pi_x \KLD{\pi_{y|x}}{\pi_{y}},
    \label{eq:mutualinfo}
\end{equation}
where $\KLD{\cdot}{\cdot}$ is the relative entropy or Kullback-Leibler divergence measured in nats (as we use natural logarithms). The mutual information only vanishes when $\pi_{y\vert x} = \pi_y \, \forall x,y$, i.e. when there is no communication.

Since the mutual information depends on the input distribution $\pi_x$, a key figure of merit is the attained mutual information in the best-case scenario, known as the channel capacity
\begin{equation}
    \mathcal{C} = \sup_{ \{ \pi_x \} } I,
    \label{eq:channel_capacity}
\end{equation}
where the supremum is taken over all input distributions. 
The noisy-channel coding theorem gives an operational interpretation of $\mathcal{C}$: when a given channel is supplemented by input encoder and output decoder functions, that encode and decode messages of length $n$, the channel capacity $\mathcal{C}$ gives the maximum average number of bits per channel use that can be reliably transmitted (i.e., with arbitrarily low probability of error), in the limit of large $n$ \cite{Cover2005Apr}. 

It is important to realize that the essence of the noisy-channel coding theorem still applies if the input distribution is not optimized. That is, for a fixed input distribution $\pi_x$, the corresponding mutual information $I$ gives the maximum average number of bits that can be reliably transmitted per channel use, for suitably designed encoder and decoder functions and in the limit of long messages.

Consider the lautum information~\cite{palomarLautumInformation2008}, defined with the reverse divergence as $L = \sum_{x} \pi_x \KLD{\pi_{y}}{\pi_{y|x}}$; it is the natural symmetrizer of the mutual information and offers an alternative information-theoretic measure of dependence. Since $L \geq 0$, the mutual information is bounded by the symmetrized $I+L$, which can be expressed as the average relative entropy between the different conditional output distributions \cite{Tasnim2024Sep}:
\begin{equation}
    I \leq I +L = \sum_{x, x'} \pi_x \pi_{x'} D( \pi_{y \vert x} \vert\vert \pi_{y \vert x'} ).
    \label{eq:mutualinfo_bound}
\end{equation}
The previous bound will be crucial in connecting with thermodynamic quantities.

%
%

\textit{Thermodynamic Cost}--- The previous description of a communication channel is purely mathematical, agnostic of the specific physical implementation of the channel. Therefore, in principle the channel specification bears no relation with physical properties such as speed and energy consumption. 

However, under a set of assumptions defining a family of communications schemes, it is possible to bound the entropy production associated with the implementation of a given channel in terms of its specification $\pi_{y|x}$ and input distribution $\pi_x$. These {\it assumptions} are: i) the output symbol $y$ is encoded in the state of a physical system whose evolution depends parametrically on the input symbol, in such a way that its stationary distribution given an input $x$ is $\pi_{y|x}$, ii) when using the channel to transmit a symbol, the input $x$ is held constant for a time $\tau$ that is long compared to the typical relaxation time of the output system, iii) when transmitting a series of symbols, the input value $x$ is switched instantaneously after having transmitted each symbol. This is not too restrictive a class, as it encompasses any channel that encodes symbols in the steady states of a physical medium and switches between them suddenly, but infrequently compared to relaxation time. Many communication channels are often operated and analyzed in this regime, as in the settled output of a logic gate or a biochemical sensing network.

Much of the physics of a communication channel is entailed in how it reacts to a change of its input. In particular, consider the relaxation process following an input switch $x\to x'$ at time $t=0$, during which the distribution $p_y(t)$ over the states of the output system evolves from the initial distribution $\pi_{y|x}$ to the final distribution $\pi_{y|x'}$.  The rate of entropy production $\dot \Sigma$ during such relaxation process can be decomposed as:
\begin{equation}
    \dot \Sigma = \dot \Sigma^\text{a} + \dot \Sigma^\text{na},
    \label{eq:ad_nonad_decomp}
\end{equation}
where $\dot \Sigma^\text{na} \equiv  - k_b d_t \KLD{p_{y\vert x'}(t)}{\pi_{y \vert x'}}$ measures the speed at which the distribution $p_{y \vert x'}(t)$ evolves towards the new steady-state, while $\dot \Sigma^a$ takes into account the contribution of probability currents that might persist even at steady-state. $k_b$ is the Boltzmann constant. Equation~\eqref{eq:ad_nonad_decomp} is known as the adiabatic/non-adiabatic decomposition of the entropy production rate and is valid for generic Markov jump processes or overdamped diffusion processes \cite{hatano2001steady, esposito2007entropy, esposito2010three, van2010three, ge2010physical} (see Appendix \ref{ap:ad_nonad_review} for a review). An important property of this decomposition is that both terms are non-negative. In particular, we have that $\dot \Sigma^\text{a} \geq 0$ and therefore
\begin{equation}
    \dot \Sigma^\text{na} = \dot \Sigma - \dot \Sigma^\text{a} \leq \dot \Sigma.
\end{equation}
Integrating this inequality from time $t=0$ up to time $t=\tau$, we obtain:
\begin{equation}
    \Sigma^\text{na}_{x \to x'} \leq
    \int_0^\tau \dot \Sigma \: dt
    \equiv 
    \Sigma_{x\to x'},
    \label{eq:rel_entropy_bound}
\end{equation}
where we defined $\Sigma_{x\to x'}$ as the total entropy produced during the relaxation process. The left-hand side is the result of integrating the non-adiabatic contribution between two steady states: $\Sigma^\text{na}_{x \to x'} \equiv k_b \KLD{\pi_{y \vert x}}{\pi_{y \vert x'}}$ since $p_y(\tau) \simeq \pi_{y|x'}$ due to assumption ii). See Appendix~\ref{ap:error_relaxation} for an analysis of the error committed by this approximation.

Notice that from Eq.~\eqref{eq:mutualinfo_bound}, $k_b I \leq \sum_{x,x'} \pi_x \pi_{x'} \Sigma^\text{na}_{x \to x'} \equiv \Sigma_\text{s}^\text{na}$, and using Eq.~\eqref{eq:rel_entropy_bound}, we obtain:
\begin{equation}
    k_b I \leq \sum_{x,x'} \pi_x \pi_{x'} \Sigma_{x\to x'} \equiv \Sigma_\text{s},
    \label{eq:main_result}
\end{equation}
where we have defined $\Sigma_\text{s}$, the average entropy produced per channel use when operating it with input distribution $\pi_x$. Note that the terms with $x=x'$ correspond to idle periods in which the symbol transmitted is the same as the previous one, and therefore the input $x$ does not change, but the system may still keep producing entropy through the adiabatic contribution. If the channel operates under detailed-balance conditions, $\Sigma_{x\to x}=0$ since $\pi_{y|x}$ is actually a thermal equilibrium state in which no entropy is produced. Channels operating in non-equilibrium conditions will have $\Sigma_{x\to x} > 0$.

The interpretation of $\Sigma_s$ in Eq. \eqref{eq:main_result} as the average entropy produced per channel use is only appropriate if the successive symbols being fed to the channel are i.i.d (i.e, they are not correlated). While this might seem an overly restrictive condition, it actually aligns with a fair characterization of communication channels. The reason is that for good encoder functions achieving $I$ transmitted bits per channel use, the statistics of short sequences of input symbols are indistinguishable from those of an i.i.d sequence \cite{Shamai1997, Polyanskiy2013}.

The bound in Eq. \eqref{eq:main_result} is our first main result. It reveals the minimal thermodynamic cost of communication under assumptions i)-iii): increasing the input-output mutual information of a channel necessarily increases the minimal amount of entropy production required for its operation. A trivial consequence is that 
\begin{equation}
k_b \mathcal{C} \leq \Sigma_\text{s}^\text{cc},   
\end{equation}
with $\Sigma_\text{s}^\text{cc}$ the entropy produced per symbol for the input distribution $\pi_x^\text{cc}$ that maximizes mutual information [cf. Eq.~\eqref{eq:channel_capacity}]. Thus, $\Sigma_\text{s}^\text{cc}$ is the entropy that must be produced to communicate at capacity.

Notice that the bound in Eq.~\eqref{eq:main_result} chains two inequalities of different nature: an information-theoretic one, Eq.~\eqref{eq:mutualinfo_bound}, and a thermodynamic one, Eq.~\eqref{eq:rel_entropy_bound}. Being independent, they saturate under separate conditions. The thermodynamic inequality saturates for detailed-balance dynamics, where $\dot\Sigma^\text{a} = 0$, so equality holds in Eq.~\eqref{eq:rel_entropy_bound}. The information-theoretic inequality, instead, has a gap given by the lautum information $L \ge 0$ that closes only in the no-communication case $\pi_{y|x}= \pi_y \: \forall x$ . Therefore, the bound never tightens for $I \gtrapprox 0$, and any channel transmitting information necessarily dissipates strictly more than $k_b I$. It is also instructive to consider the low-information regime around the no-communication case, defined formally as $\delta(\pi_{y|x}, \pi_{y|x'}) \leq h \to 0$, where $\delta(\cdot, \cdot)$ is the total variation distance. Since $I=L$ to the lowest (quadratic) order in $h$, we have that $k_b I\simeq \Sigma^\text{na}_s/2$ in the low-information regime (see App.~\ref{ap:saturation_bound}). Therefore, for marginally communicating channels operating in detailed balance conditions (where $\Sigma^\text{a}_s \equiv \Sigma_s - \Sigma^\text{na}_s = 0$), mutual information and entropy production are directly related by a factor 2.

It is worth emphasizing the distinct physical status of the adiabatic and non-adiabatic contributions to the total entropy production. For non-equilibrium channels, the adiabatic entropy production is extensive in the holding time $\tau$, since $\dot \Sigma^\text{a}$ persists even after relaxation. Thus, holding each input for longer necessarily dissipates more and renders the channel arbitrarily inefficient as $\tau\to\infty$. The non-adiabatic contribution  $\Sigma_\text{s}^\text{na}$, by contrast, is intensive in $\tau$: it is produced only during the transient relaxation that follows an input switch and saturates to $\sum_{x,x'}\pi_x \pi_{x'} \KLD{\pi_{y|x}}{\pi_{y|x'}}$ regardless of how long the symbol is subsequently held. It only depends on channel specification and not on dynamics, thus isolating the more fundamental, time-independent cost intrinsic to communication, bounding on its own the mutual information. The total entropy production $\Sigma_\text{s}$ remains the physical quantity of ultimate relevance, as it measures the energy dissipated while operating the channel.

\textit{Thermodynamic efficiency and its optimization}--- The bound $k_b I\le\Sigma_\text{s}$ invites a natural figure of merit, the thermodynamic efficiency of a channel,
\begin{equation}
    \eta \equiv \frac{k_b I}{\Sigma_\text{s}} \leq 1,
\end{equation}
which quantifies the information transmitted per unit of entropy produced. Its tighter non-adiabatic counterpart $\eta^\text{na}\equiv k_b I/\Sigma_\text{s}^\text{na}\le 1$ captures only the time-independent cost that saturates at $1/2$ in the low-information regime.

Unlike the channel capacity, whose value is operationally approached through coding, the upper bound $\eta=1$ is reached by no useful channel: as discussed above, the information inequality tightens only in the no-communication limit. The pertinent question is therefore not whether the bound is saturated, but which input distribution lets the channel operate as efficiently as possible. This defines an optimization where rather than maximizing the transmitted information regardless of its thermodynamic cost [cf. Eq.~\eqref{eq:channel_capacity}], we maximize the information relative to that cost. As we show, the two optimizations generally select different inputs.

Using the method of Lagrange multipliers to optimize the mutual information in Eq. \eqref{eq:mutualinfo} with respect to the input distribution $\pi_x$ we find that, if the optimal distribution $\pi_x^\text{cc}$ is inside the probability simplex, it must satisfy:
\begin{equation}
    \mathcal{C} = \KLD{\pi_{y|x}}{\pi^\text{cc}_y} \quad \forall x,
\end{equation}
with $\pi_y^\text{cc} = \sum_x \pi_{y|x} \pi^\text{cc}_x$.
Therefore, the optimal input distribution is such that the unconditional distribution $\pi_y^\text{cc}$ is `equidistant' from all conditional distributions $\pi_{y|x}$ (loosely considering the relative entropy as a distance). That common distance is itself the channel capacity.

An analogous calculation shows that the input distribution $\pi_x^\text{me}$ maximizing the efficiency, if interior to the simplex, must satisfy
\begin{equation}
    \eta = \frac{k_b \KLD{\pi_{y|x}}{\pi^\text{me}_{y}}}{\partial_{\pi_x} \Sigma_\text{s} - \Sigma_\text{s}} \quad \forall x,
    \label{eq:optimal_efficiency}
\end{equation}
where $\partial_{\pi_x} \Sigma_\text{s} = \sum_{x'} \pi_{x'}( \Sigma_{x \to x'} + \Sigma_{x' \to x})$ and all  quantities are evaluated at $\pi_x^\text{me}$. If for a given implementation of a channel the entropy production is the same for all possible transitions on one hand, and for all idle periods on the other hand, i.e., if $\Sigma_{x \to x'} = \Sigma_T$ for all $x\neq x'$ and $\Sigma_{x \to x} = \Sigma_I$ for all $x$, then we have $\partial_{\pi_x} \Sigma_\text{s} = 2 (1-\pi_x)\Sigma_T + 2\pi_x \Sigma_I$ and $\Sigma_\text{s} = \Sigma_T + (\Sigma_I - \Sigma_T)\sum_{x} \pi_x^2$. Thus, we see that the denominator in Eq. \eqref{eq:optimal_efficiency}, $\partial_{\pi_x} \Sigma_\text{s} - \Sigma_\text{s}$, only becomes independent of $x$ in the trivial case $\Sigma_T = \Sigma_I$, where the entropy production of the channel is completely independent of the input distribution. Only in that case is the optimization of efficiency equivalent to the optimization of the mutual information. If instead $\Sigma_T > \Sigma_I$, the optimization of the efficiency will favor input distributions that are more pure, in order to minimize the probability of transitions where the input changes. This happens even in the case where the channel and its physical implementation are symmetric with respect to all inputs and transitions, as we now exemplify.

\textit{Binary symmetric channel}--- We now study the consequences of optimizing the thermodynamic efficiency in the simplest example of a communication channel: the binary symmetric channel (BSC). In this case the input and output alphabets are $\mathcal{X} = \mathcal{Y}= \{0,1\}$, and the channel is defined by $\pi_{y|x} = 1-\epsilon$ if $x=y$ and $\pi_{y|x} = \epsilon$ if $x\neq y$. Thus, the channel transmits both inputs with the same error probability $\epsilon$. It follows that the input distribution achieving channel capacity is $\pi_x = 1/2$ and that $\mathcal{C}_\text{BSC} = \ln 2 - H_b(\epsilon)$, where $H_b(\epsilon) \equiv -\epsilon \ln(\epsilon) -(1-\epsilon)\ln(1-\epsilon)$ is the binary entropy function.

To study the thermodynamic cost, we consider a physical implementation that respects the symmetry in the channel specification. Thus, we assume that both possible input transitions have the same entropy production, $\Sigma_{0\to1}=\Sigma_{1\to 0} = \Sigma_T$, and the same idle periods, $\Sigma_{0 \to 0} = \Sigma_{1\to 1} = \Sigma_I$. The input distribution $\pi_{x}^\text{me}$ maximizing the efficiency $\eta$ can then be computed as a function of $\epsilon$ and $r\equiv \Sigma_T/\Sigma_I$ for fixed $\Sigma_I$.

\begin{figure}[t]
    \includegraphics[width=\linewidth]{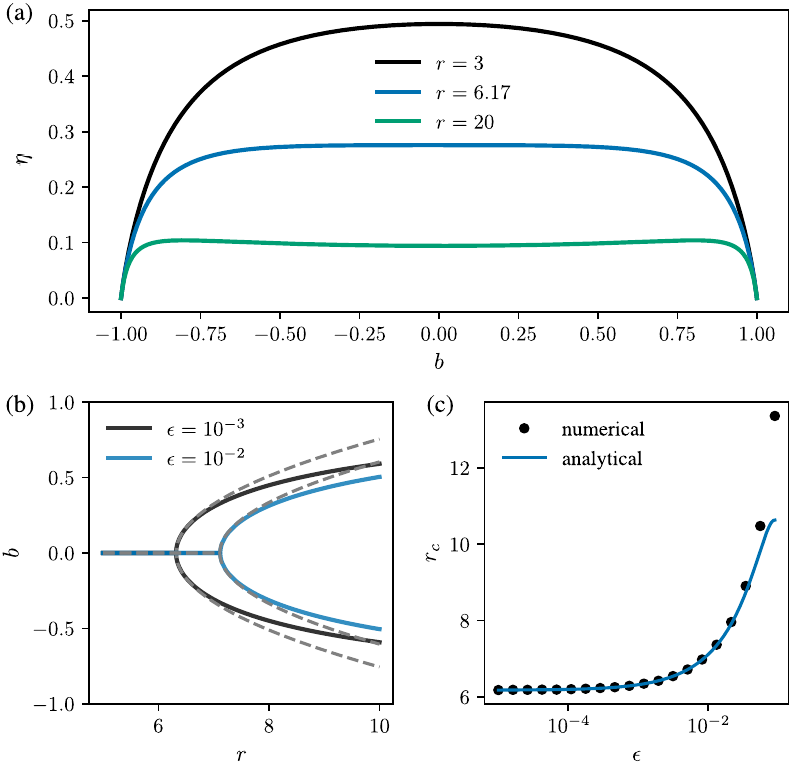}
    \caption{\textbf{Symmetry breaking in the Binary Symmetric Channel at maximum efficiency.} (a) Efficiency as a function of input bias for $\epsilon=10^{-3}$ and different values of $r = \Sigma_T/\Sigma_I$ ($\Sigma_I/k_b = 1$). (b) Input bias maximizing efficiency as a function of $r$ for different values of $\epsilon$. (c) Critical point $r_c$ as a function of $\epsilon$ (see Eq. \eqref{eq:crit_r}).}
    \label{fig:bsc}
\end{figure}

Despite the simplicity of the problem, we show below that the behavior of $\pi_x^\text{me}$ is surprisingly rich. In order to see this, we parametrize the input distribution as $\pi_{0/1} = (1\mp b)/2$, as a function of a bias parameter $b$. In Fig. \ref{fig:bsc}-(a) we show the efficiency $\eta$ as a function the bias $b$ for different values of $r$. We see that when $r$ is below a critical threshold, the input distribution maximizing efficiency is the symmetric distribution $\pi_0 = \pi_1 = 1/2$, as the optimal value of the bias is $b=0$. However, above a given value of $r$ this symmetry is broken, and $b=0$ ceases to be a local optimum through a pitchfork bifurcation leading to optimal biases $\pm b^*$ with $b^* > 0$.

This symmetry breaking behavior is more evident in Fig. \ref{fig:bsc}-(b), where we plot the optimal values of $b$ as a function of $r$, for different values of $\epsilon$. We see that the symmetry in the input distribution is broken when $r$ is above a critical value $r_c$, which increases with increasing error probability $\epsilon$. In fact, in the limit of small $\epsilon$, the critical point can be found to be:
\begin{equation}
    r_c = \frac{\ln2 + \epsilon(\ln(\epsilon)-3) + 1/2 }{\ln(2) + \epsilon(\ln(\epsilon)+1) - 1/2}. 
    \label{eq:crit_r}
\end{equation}
In Fig. \ref{fig:bsc}-(c) we compare the previous expression with the numerically obtained critical point. Finally, a simple calculation shows that for $r\gtrsim r_c$ the optimal value of the input bias is $b \simeq \sqrt{12\ln(2) (r-r_c)}/(r_c+1)$
(this is indicated by dashed lines in Fig. \ref{fig:bsc}-(b)).

Symmetry breaking is favored by less dissipative channels. A large idle cost $\Sigma_I$ keeps $r<r_c$ and the input symmetric, while near equilibrium ($\Sigma_I\to0$, $r\to\infty$) the efficiency-optimal input is always biased toward a low-entropy distribution that minimizes costly switches.

\begin{figure}
    \includegraphics[scale=.7]{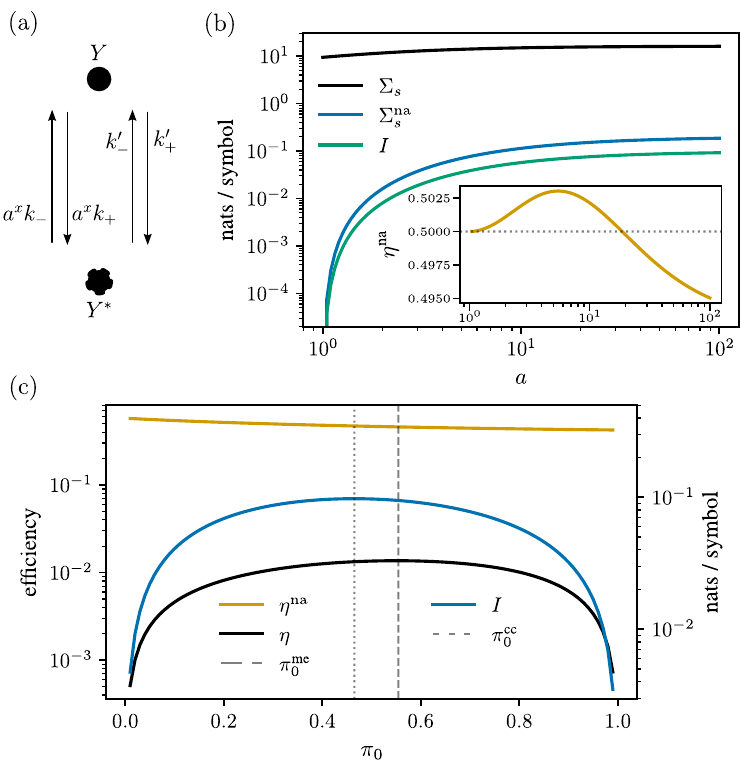}
    \caption{\textbf{Thermodynamic limits of the sensing channel.} (a) Scheme of the possible transitions in the simple sensing model. (b) Mutual information $I$, its non-adiabatic bound $\Sigma^\text{na}_\text{s}$, and the non-adiabatic efficiency $\eta^\text{na} = k_b I/\Sigma_\text{s}^\text{na}$ (inset) versus the rescaling factor $a$ at $\pi_0=0.3$. (c) Mutual information $I$ and efficiencies $\eta$ and $\eta^\text{na}$ versus input probability $\pi_0$ at $a=50$. The total entropy production $\Sigma_s$ entering $\eta=k_b I/\Sigma_s$ is computed for a holding time $\tau$ equal to 5 times the slowest relaxation time of the output (see Appendix \ref{app:sensing}). 
    The dashed lines indicate the optimal input probabilities $\pi_0^\text{me}$ and $\pi_0^\text{cc}$. In all cases we took $k_+ = k'_-  = 1.0$, $k_- = k'_+ = 0.1$ and $k_b=T=1$.}
    \label{fig:sensing}
\end{figure}

\textit{Sensing model}---Chemotaxis is essential to the survival of unicellular organisms such as bacteria, enabling them to navigate chemical gradients toward nutrients and away from harmful substances~\cite{bergPhysicsChemoreception1977}. This is achieved through membrane-bound receptors that stochastically bind and unbind environmental ligands. These binding events modulate the phosphorylation activity of downstream signaling proteins, that as a result carry information about the local ligand concentration. Many works have since investigated the fundamental limits of this sensing process~\cite{bergPhysicsChemoreception1977, bialekPhysicalLimitsBiochemical2005, moraLimitsSensingTemporal2010a, governFundamentalLimitsSensing2012, mehtaEnergeticCostsCellular2012, baratoInformationtheoreticThermodynamicEntropy2013a, harveyUniversalEnergyaccuracyTradeoffs2023c}, including their energetics.

For illustrative purposes, we consider a minimal model of one receptor and a single downstream protein~\cite{baratoInformationtheoreticThermodynamicEntropy2013a}, which captures the basic properties of sensing. 
We consider that the receptor has a stationary probability $\pi_0$ of being empty and $\pi_1=1-\pi_0$ of being bound to a ligand. Binding/unbinding events happen a rates proportional to $\tau^{-1}$, setting the timescale of input switches. Thus trajectories of the receptor state constitute the input signal, with alphabet $\mathcal{X} = \{ 0,1 \}$. Meanwhile, a downstream protein can be in an inactive (dephosphorylated) state $\text{Y}$ or an active (phosphorylated) state $\text{Y}^*$, which constitutes the output alphabet $\mathcal{Y} = \{ \text{Y}, \text{Y}^* \}$. The protein switches between these two states through two possible reversible mechanisms: a ligand-dependent phosphorylation reaction $\text{Y} + \text{ATP} \xrightleftharpoons[]{} \text{Y}^* + \text{ADP}$ and a ligand-independent dephosphorylation reaction $\text{Y}^* \xrightleftharpoons[]{} \text{Y} + \text{Pi}$. The former is modulated by the receptor state $x$, in such a way that the phosphorylation rates $k_\pm$ are rescaled by a factor $a \ge 1$ when the receptor is bound ($x=1$). The latter induces spontaneous transitions at rates $k'_\pm$ that are independent of the receptor state. The two reactions form a cycle fueled by ATP hydrolysis with affinity $\Delta\mu = k_b T \ln (k_+ k'_- / k_- k'_+ )$, which drives the system out of equilibrium whenever $\Delta\mu \neq 0$. See Fig. \ref{fig:sensing}-(a) for a schematic representation of the model.


This model satisfies assumptions i)–iii): the protein encodes the receptor state in its ATP-driven steady state, held for a time $\tau$ long compared to the output relaxation (see App. \ref{app:sensing}). Figure~\ref{fig:sensing}(b) confirms the central bound $k_b I \leq \Sigma^\text{na}_\text{s} \leq \Sigma_\text{s}$ across the whole coupling range. The two gaps have distinct physical origins: $\Sigma^\text{na}_\text{s}-k_b I$ is the information-theoretic slack (the lautum information), while $\Sigma_\text{s}-\Sigma^\text{na}_\text{s}$ is the adiabatic cost of the phosphorylation cycle, i.e. the energy dissipated by the cell to keep the readout responsive even while the receptor state is held fixed. In the weak-sensing limit $a\to1$, the non-adiabatic efficiency saturates at $\eta^\text{na} \to 1/2$ [inset], where sensing dissipates at least twice the information it acquires. Figure~\ref{fig:sensing}(c) shows the values of efficiencies $\eta$ and $\eta^\text{na}$, as well as the mutual information $I$, with respect to the probability of an empty receptor $\pi_0$. The $\pi_0$ that maximizes efficiency, $\pi_0^\text{me}$, does not coincide with the capacity-achieving one, $\pi_0^\text{cc}$. 
Hence a receptor operating with the most information per channel use is not transmitting the most information per unit dissipation. We see that energetic efficiency favors a biased receptor occupancy that lowers the rate of costly binding/unbinding switches.

\emph{Discussion.}--- It is important to distinguish our thermodynamic analysis of communication from previous studies about the thermodynamics of \emph{information rates}~\cite{Tostevin2009May, baratoInformationtheoreticThermodynamicEntropy2013a} and \emph{information flows}~\cite{horowitzThermodynamicsContinuousInformation2014}. These are related but different concepts, better explained in the context of bipartite systems. On one hand, the information rate measures how the mutual information between entire trajectories of different parts of a system increases with time. This quantity has been shown to \emph{not} be bounded by the entropy production rate \cite{baratoInformationtheoreticThermodynamicEntropy2013a}, and should be distinguished from the \emph{static} mutual information considered here as it takes into account arbitrarily delayed temporal correlations. On the other hand, one can consider the mutual information $I_t$ between two parts of a bipartite system at a given time $t$. The temporal evolution of this quantity accepts a decomposition into a balance equation of the form $\dot I_t= \dot I_t^X + \dot I_t^Y$, where each term in the right-hand side accounts for the contribution of the local dynamics of each part. These contributions are known as information flows; they are bounded in terms or energy, matter, and entropy flows through generalized second-law-like inequalities, and have been instrumental in understanding the thermodynamics of feedback control and information engines \cite{horowitzThermodynamicsContinuousInformation2014, hartichStochasticThermodynamicsBipartite2014}. 

Our analysis is complementary to those previous studies, as it focuses on the thermodynamic cost of achieving a particular level of mutual information between the input and output of a communication channel, under the time-scale separation conditions i-iii), rather than on its flow between subsystems or the rate of change of trajectory-level mutual information. The treatment presented here better aligns with the concept of communication channel as conceived in information theory. It properly characterizes the thermodynamic cost of their sequential operation, and can be directly applied to real systems like electronic gates or chemical networks.

A related analysis of communication channels based on mismatch
cost was presented in Ref.~\cite{yadavMinimalThermodynamicCost2025a}.
There, communication is modeled as a cycle consisting of an
independent overwriting of the input followed by a copying operation.
The mismatch cost of overwriting is bounded below by the
mutual information $I$, while for specific choices of the
thermodynamic priors, the mismatch costs of overwriting and
copying become $I$ and $L$, respectively.
Thus, the combination $I+L$ appearing in our result also admits
an interpretation as the sum of mismatch costs in that protocol.
The two approaches, however, employ different 
descriptions: here the input is treated as an external control
parameter, and the entropy production is evaluated from the
relaxation of the output conditioned on successive input values.
See App. \ref{Missmatch} for details.

Finally, a superficially similar symmetry breaking behavior has been reported for thermodynamically optimal copying~\cite{mulder2025exploiting}. There, minimizing the finite-time work to copy an unbiased data bit at fixed generated mutual information drives the optimal initial memory distribution away from the symmetric point, together with an asymmetry in the copying accuracies. Despite the similarities, the mechanism differs from ours: the distribution being optimized is a protocol parameter rather than the input statistics, and the information transmitted is held fixed rather than traded against dissipation.

More broadly, information-theoretic and thermodynamic quantities have been linked in a variety of settings: the cost of measurement and erasure~\cite{sagawaMinimalEnergyCost2009a}, the value of predictive information~\cite{stillThermodynamicsPrediction2012a}, sensing and transduction in living systems~\cite{tkacikInformationProcessingLiving2016a, brittainWhatWeLearn2017, nicolettiTuningTransductionHidden2024}, interacting or partially observed subsystems~\cite{chetriteInformationThermodynamicsInteracting2019, leightonInferringSubsystemEfficiencies2023a, leightonFlowEnergyInformation2025}, finite-time bounds on speed, error, and dissipation~\cite{kamijimaFinitetimeThermodynamicBounds2025, vanvuTimeCostErrorTradeOffRelation2025}, exact information--dissipation identities for continuous-state dynamics~\cite{choExactIdentityLinking2026}, mismatch-cost decompositions of stochastic maps~\cite{yadavMinimalThermodynamicCost2025a}, and error--dissipation trade-offs for an abrupt switch between two stationary states~\cite{Falasco_JPhysA_2022}, of which Eq.~\eqref{eq:main_result} is the many-symbol, arbitrary-input extension.

\emph{Acknowledgments}---App. \ref{Missmatch} resulted from discussions with the authors of \cite{yadavMinimalThermodynamicCost2025a}.
PH is supported by the project INTER/FNRS/20/15074473 funded by F.R.S.-FNRS (Belgium) and FNR (Luxembourg), and by the European Union (ERC-SuperStoc-101117322).
ME is funded by the Fonds National de la Recherche-FNR, Luxembourg: project NEQPHASETRANS (C24/MS/18933049).

\clearpage
\appendix

\onecolumngrid
\begin{center}
    \textbf{End Matter}
\end{center}
\vspace{1em}
\twocolumngrid

\section{Review of Adiabatic/Non-Adiabatic decomposition}
\label{ap:ad_nonad_review}

Consider a continuous-time Markov process over output states $y\in\mathcal{Y}$ with transition rates $w_{yy'}$ conditioned on a fixed input $x'$. The probability distribution $p_y(t)$ evolves according to
\begin{equation}
    d_t p_y(t) = \sum_{y'} \bigl[ w_{yy'} p_{y'}(t) - w_{y' y} p_y(t) \bigr].
\end{equation}
For a given input $x'$, the stationary distribution satisfies $\sum_{y'} [w_{yy'}\pi_{y'|x'} - w_{y'y}\pi_{y|x'}]=0$ and is denoted $\pi_{y|x'}$. If in a particular system the transition $y' \to y$ can occur via two or more independent physical mechanisms, we assume that the total transition rate can be decomposed as $w_{yy'}=\sum_\nu w_{yy'}^\nu$, where $w_{yy'}^\nu$ is the rate corresponding to mechanism $\nu$.   

The entropy production rate for this process can be written in the familiar Schnakenberg form
\begin{equation}
    \dot \Sigma = \frac{k_b}{2} \sum_\nu \sum_{y,y'} J^\nu_{yy'}(t) \ln \frac{w^\nu_{yy'} p_{y'}(t)}{w^\nu_{y'y} p_y(t)},
\end{equation}
with the instantaneous probability current $J^\nu_{yy'}(t) = w^\nu_{yy'} p_{y'}(t) - w^\nu_{y'y} p_y(t)$.
This rate is naturally decomposed into an adiabatic part,
\begin{equation}
    \dot \Sigma^\text{a} = \frac{k_b}{2} \sum_\nu \sum_{y,y'} J^\nu_{yy'}(t) \ln \frac{w^\nu_{yy'} \pi_{y'|x'}}{w^\nu_{y'y} \pi_{y|x'}},
\end{equation}
and a non-adiabatic part,
\begin{equation}
    \dot \Sigma^\text{na} = -k_b \frac{d}{dt} D\bigl(p_y(t) \big\| \pi_{y\mid x'}\bigr),
\end{equation}
where $D(p\|q) = \sum_y p_y \ln(p_y/q_y)$ is the Kullback-Leibler divergence. The total entropy production rate is the sum
\begin{equation}
    \dot \Sigma = \dot \Sigma^\text{a} + \dot \Sigma^\text{na}.
\end{equation}

Both contributions are non-negative for generic Markov jump processes and overdamped diffusion dynamics \cite{hatano2001steady, esposito2010three, van2010three, ge2010physical}. The non-adiabatic term measures the relaxation of $p_y(t)$ toward the new steady state $\pi_{y|x'}$, while the adiabatic term measures the steady-state dissipation associated with currents that persist even when $p_y(t)=\pi_{y|x'}$.

If the input switches from $x$ to $x'$ at time $t=0$, then $p_y(0)=\pi_{y|x}$ and the integrated non-adiabatic entropy production over the subsequent relaxation satisfies
\begin{equation}
    \int_0^\tau \dot \Sigma^\text{na}(t) \,dt = k_b D(\pi_{y|x} \| \pi_{y|x'}) - k_b D(p(\tau) \| \pi_{y|x'}),
\end{equation}
so that for $\tau$ much larger than the relaxation time the boundary term becomes negligible and
\begin{equation}
    \Sigma^\text{na}_{x\to x'} \simeq k_b D(\pi_{y|x} \| \pi_{y|x'}).
\end{equation}
This is the key relation used in the main text to bound the mutual information by the entropy produced during input switching.

\section{Relaxation time approximation}
\label{ap:error_relaxation}

Under assumption ii), each input symbol $x$ is held to a time $\tau$ larger than the typical relaxation time. This means that the integral involved in $\Sigma_{x \to x'}$ will approximately start in $\pi_{y \vert x}$ and end in $\pi_{y \vert x'}$.

With fixed $x'$, consider that at time $\tau$ the output distribution is close to its next steady-state as  $p_{y \vert x'} (\tau) = \pi_{y \vert x'} + h f_{x'}(y)$, with $\sum_y f_{x'}(y) = 0$ ensuring probability conservation and $h$ being a small parameter. The non-adiabatic contribution that bounds the mutual information evaluates as
\begin{equation}
	\Sigma^\text{na}_{x \to x'}/k_b = \KLD{ \pi_{y \vert x} }{ \pi_{y \vert x'} } - h^2 \sum_y \frac{f_{x'}^2(y)}{ 2 \pi_{y \vert x'} } + \mathcal{O} [h^3 ].
\end{equation}
The bound on the mutual information
\begin{equation}
    I \leq \Sigma^\text{na}_\text{s}/k_b + h^2 \sum_{x',y} \frac{\pi_{x'} f_{x'}^2(y)}{2 \pi_{y \vert x'}} + \mathcal{O}[h^3],
\end{equation}
will acquire a correction that is only of second order in $h$. It tightens quadratically with assumption ii) and is thus robust to small deviations.

\section{Low-mutual information regime}
\label{ap:saturation_bound}

When $x$ and $y$ are independent, both the mutual information and its bound in Eq.~\eqref{eq:mutualinfo_bound} vanish. We perturb the conditional distribution around this independence regime, where input and output are weakly coupled:
\begin{equation}
    \pi_{y \vert x} = \pi_y + h g_x(y) + \mathcal{O} [h^2].
\end{equation}
Normalization of $\pi_{y|x}$ imposes $\sum_y g_x (y) = 0$. In the low-information regime, both the mutual and lautum information have the same scaling to leading order in $h$:
\begin{equation}\label{eq:I_lowinfo}
    I = L = h^2 \sum_{xy} \frac{\pi_x g_x^2 (y)}{2 \pi_y} + \mathcal{O} [h^3].
\end{equation}
Since the bound in Eq.~\eqref{eq:mutualinfo_bound} relates $I$ and $I+L$, it will saturate with a factor of 2 in the low-information regime:
\begin{equation}
	k_b I \to \frac{k_b}{2} \sum_{x,x'} \pi_x \pi_{x'} D( \pi_{y\vert x} \vert\vert \pi_{y \vert x'}) = \frac{1}{2} \: \Sigma_s^\text{na}.
\end{equation}

\section{Calculations for the BSC}

We gather here some expressions behind the analysis of the binary symmetric channel. We first compute the ingredients going into Eq. \eqref{eq:mutualinfo} for the mutual information:
\begin{equation}
\begin{split}
D_0 \equiv D(\pi_{y|0}||\pi_y) = -\ln(\pi_0) - \epsilon &\ln(1+\frac{\pi_1}{\pi_0} \frac{1-\epsilon}{\epsilon})  \\
- (1-\epsilon) &\ln\left(1+\frac{\pi_1}{\pi_0} \frac{\epsilon}{1-\epsilon}\right).
\end{split}
\end{equation}
An analogous expression for $D_1 \equiv D(\pi_{y|1}|\pi_y)$ follows by just interchanging $\pi_0 \leftrightarrow \pi_1$. To linear order in $\epsilon$, the mutual information can then be expanded as:
\begin{equation}
    I = H(x) + \epsilon(\ln(\epsilon) - 1) + \epsilon (\pi_1 - \pi_0) \ln(\pi_1/\pi_0)+ \mathcal{O}(\epsilon^2).
\end{equation}

Now we just parametrize the input distribution as $\pi_{0} = (1 - b)/2$ and $\pi_{1} = (1 + b)/2$, with $-1\leq b \leq 1$, and approximate $H(x)$ to lower order in $b$, obtaining
\begin{equation}
    I = \ln(2) + \epsilon(\ln(\epsilon)-1) - b^2(1/2-2\epsilon) + \mathcal{O}(\epsilon^2).
\end{equation}
The second ingredient needed is the entropy production. It reads
\begin{equation}
    \Sigma_\text{s} = \frac{1}{2}(\Sigma_I + \Sigma_T) + \frac{1}{2} b^2 (\Sigma_I - \Sigma_T)
\end{equation}
Thus, to linear order in $\epsilon$ and second order in $b$, the efficiency is:
\begin{equation}
    \eta = \frac{\ln(2) + \epsilon(\ln(\epsilon)-1) - b^2(1/2-2\epsilon)}{\frac{1}{2k_b}(\Sigma_I + \Sigma_T) + \frac{1}{2k_b} b^2 (\Sigma_I - \Sigma_T)}
\end{equation}
Now we consider the problem of optimizing $\eta$ with respect to the bias $b$. It turns out the unbiased symmetric point $b=0$ is a local optimum whenever 
\begin{equation}
    \frac{r-1}{r+1} \leq \frac{1/2 - 2 \epsilon}{\ln(2) + \epsilon (\ln(\epsilon)-1)},
\end{equation}
where we defined $r \equiv \Sigma_T/\Sigma_I$. Then, a symmetric input distribution is optimal if:
\begin{equation}
    r \leq r_c \equiv \frac{\ln(2) + \epsilon(\ln(\epsilon)-3) + 1/2 }{\ln(2) + \epsilon(\ln(\epsilon)+1) - 1/2}.
\end{equation}
The critical $r$ takes the value $r_c \simeq 6.17$ in the limit $\epsilon \to 0$. 

\section{Calculations for the chemical sensing example}
\label{app:sensing}

With fixed input $x$, the steady-state probability $P(Y^*|x)$ for the downstream protein $Y$ to be phosphorylated is given by:
$\pi_{y=1|x} = \tau_x (k'_+ + a^x k_+ )$, where $\tau_x \equiv (k'_- + k'_+ + a^x (k_- + k_+) )^{-1}$ is the timescale at which the system relaxes to the steady state. If the typical timescale between input switches is $\tau \gg \tau_x \forall x$, we have $\Sigma_{x\to x'} = \Sigma_{x\to x'}^\text{a} + \Sigma_{x\to x'}^\text{na} $, with $x' =  1-x$ and
\begin{equation}
\begin{split}
    \Sigma^\text{na}_{x\to x'} & \simeq k_b \KLD{\pi_{y \vert x}}{\pi_{y \vert x'}}\\
    & = k_b \pi_{1|x} \ln\left(\frac{\pi_{1|x}}{\pi_{1|x'}}\right) + k_b(1-\pi_{1|x}) \ln\left(\frac{1-\pi_{1|x}}{1-\pi_{1|x'}}\right),\\
    \Sigma^\text{a}_{x\to x'} & \simeq \tau \dot \Sigma_{x'},
\end{split}
\end{equation}
where $\dot \Sigma_{x'}$ is the steady-state entropy production rate for fixed input $x'$, given by:
\begin{equation}
\begin{split}
    \dot \Sigma_{x'} &= \underbrace{[\pi_{0|x'} \: a^{x'} k_+ - \pi_{1|x'} \: a^{x'} k_-]}_{J^a} k_b \ln(k_+/k_-) \\
                     &+ \underbrace{[\pi_{0|x'} \: k'_+ - \pi_{1|x'} \: k'_-]}_{J^b} k_b \ln(k'_+/k'_-) \geq 0.
\end{split}
\end{equation}
Note that, due to the vanishing of the net steady-state current $J = J^a + J^b=0$, $\dot \Sigma_{x'}=0$ when  $\Delta\mu \equiv k_b T \ln(k_+k'_-/k_-k'_+) = 0$.

\section{Connection with the mismatch-cost approach}
\label{Missmatch}

In Ref.~\cite{yadavMinimalThermodynamicCost2025a},
communication is represented by two successive stochastic
maps: an overwriting operation (a) and a copying operation (b),
defined by
\begin{align}
G_a(x',y'|x,y) &= \pi_{x'}\delta_{y',y},\\
G_b(x',y'|x,y) &= \delta_{x',x}\pi_{y'|x}.
\end{align}
The corresponding transformations of the joint distribution
are
\begin{equation}
\pi_x\pi_{y|x}
\xrightarrow{\;a\;}
\pi_x\pi_y
\xrightarrow{\;b\;}
\pi_x\pi_{y|x}.
\end{equation}

The mismatch cost associated with each operation is
$\mathrm{MC}=D(p\|q)-D(Gp\|Gq)$, where $q$ is the
initial distribution minimizing the entropy production
of the corresponding physical process.

For overwriting, the operation acts on $x$ independently of
$y$, including its entropy flow. This implies that the
minimizing prior factorizes as $q_{xy}^a=q_x^a q_y^a$.
The corresponding mismatch cost is
\begin{align}
\mathrm{MC}_a
&=D(\pi_x\pi_{y|x}\|q_x^a q_y^a)
 -D(\pi_x\pi_y\|\pi_x q_y^a)
\nonumber\\
&=I+D(\pi_x\|q_x^a).
\end{align}

For copying, the output dynamics depend on the input,
so the prior may have the conditional form
$q_{xy}^b=q_x^b q_{y|x}^b$. One obtains
\begin{align}
\mathrm{MC}_b
&=D(\pi_x\pi_y\|q_x^b q_{y|x}^b)
 -D(\pi_x\pi_{y|x}\|q_x^b\pi_{y|x})
\nonumber\\
&=\sum_x\pi_xD(\pi_y\|q_{y|x}^b).
\end{align}

For the particular choices
$q_x^a=\pi_x$ and $q_{y|x}^b=\pi_{y|x}$,
the two contributions reduce to
\begin{equation}
\mathrm{MC}_a=I,\qquad
\mathrm{MC}_b=L,
\end{equation}
and hence
\begin{equation}
\mathrm{MC}_a+\mathrm{MC}_b=I+L.
\end{equation}

The second choice is justified, for example, when copying
is implemented through detailed-balanced relaxation toward
$\pi_{y|x}$, which minimizes the entropy production for
each fixed input. For general nonequilibrium dynamics,
the entropy-production-minimizing prior need not coincide
with the stationary distribution.

This provides a mismatch-cost interpretation of the same
information-theoretic combination appearing in our
nonadiabatic entropy production. In our approach, however,
the input is treated as an externally switched control
parameter, and the relaxation is evaluated conditional
on both the previous and the new input. No separate
thermodynamic cost is assigned to overwriting the input.

\bibliographystyle{apsrev4-2}
\bibliography{refs}

\end{document}